# Experimental study of atmospheric pressure single-pulse nanosecond discharge in pin-to-pin configuration

Xingxing Wang, Adam Patel, Sally Bane, Alexey Shashurin

Purdue University, 701 W Stadium Ave, West Lafayette, IN 47907, United States of America

## Abstract

We present an experimental study of nanosecond high-voltage discharges in a pin-to-pin electrode configuration at atmospheric conditions operating in single-pulse mode (no memory effects). Discharge parameters were measured using microwave Rayleigh scattering, laser Rayleigh scattering, optical emission spectroscopy enhanced with a nanosecond probing pulse, and fast photography. Spark and corona discharge regimes were studied for electrode gap sizes 2-10 mm and discharge pulse duration of 90 ns. The spark regime was observed for gaps < 6 mm using discharge pulse energies of 0.6-1 mJ per mm of the gap length. Higher electron number densities, total electron number per gap length, discharge currents, and gas temperatures were observed for smaller electrode gaps and larger pulse energies, reaching maximal values of about $7.5 \times 10^{15}$ cm$^{-3}$, $3.5 \times 10^{11}$ electrons per mm, 22 A, and 4,000 K (at 10 µs after the discharge), respectively, for a 2 mm gap and 1 mJ/mm discharge pulse energy. Initial breakdown was followed by a secondary breakdown occurring about 30-70 ns later and was associated with ignition of a cathode spot and transition to cathodic arc. A majority of the discharge pulse energy was deposited into the gas before the secondary breakdown (85-89%). The electron number density after the ns-discharge pulse decayed with a characteristic time scale of 150 ns governed by dissociative recombination and electron attachment to oxygen mechanisms. For the corona regime, substantially lower pulse energies (~0.1 mJ/mm), peak conduction current (1-2 A), and electron numbers (3-5×10$^{10}$ electrons per mm), and gas temperatures (360 K) were observed.

# Introduction

Nanosecond-scale pulsed plasma discharges (ns-discharges) have recently been proven valuable over a wide range of applications in the fields of combustion, aerodynamics, medicine, and many others. The nanosecond-scale rise time of the voltage pulse up to high-kV values enables the local electric field to increase rapidly before breakdown, energizing electrons to very high energy levels.[1] This enables production of a variety of reactive species (ozone, $NO_x$, OH, O, etc.). During the relaxation of excited and ionized species, a large amount heat can be released leading to so-called "fast" gas heating.[2,3] Additionally, fast gas heating generates strong pressure gradients, resulting in a shock wave and induced vorticity that can be further used to fulfill a particular aerodynamic purpose.

Ns-discharges have been widely applied in the field of combustion for multiple purposes, including sustaining fuel-lean combustion by providing a substantial amount of active species (O, H, OH, etc.), and providing additional gas heating to accelerate chemical reactions[4–6], increasing the flame speed, and mitigating flame instabilities.[7–12] In the field of aerodynamics, ns-discharges have been utilized as flow control actuators in the form of dielectric barrier discharges (DBDs) on the surface of an airfoil. The boundary layer is modified through gas heating which leads to a delay of the flow separation, and ultimately results in a reduction of drag.[13,14] Other aerodynamic applications of ns-discharges include shock wave modification and instability-induced flow manipulation.[15–17] Furthermore, ns-discharges operating at atmospheric conditions have also found merit in a diverse range of fields such as medicine, nanotechnology, material processing and sterilization.[18]

A comprehensive study of the properties of ns-discharges is critical for tailoring the plasma to a particular application. A logical first step in conducting that study would be to analyze a single-pulse ns-discharge before investigating the more complicated regime of nanosecond repetitively pulsed (NRP) discharges where a memory effect between the adjacent pulses is present, i.e., when perturbations induced by an individual HV pulse would not decay fully before the subsequent pulse. However, the vast majority of studies on ns-discharges in the literature were conducted using NRP discharges with high repetition frequencies (~10 kHz). Previous investigations of NRP plasmas have used multiple diagnostic techniques including optical emission spectroscopy (OES), Stark broadening, laser Thomson/Rayleigh scattering, laser induced fluorescence (LIF), microwave Rayleigh scattering, electrical measurements (voltage, current, resistance of the plasma column), etc.[19–26] A comprehensive study of the single-pulse ns-discharge is still lacking and would provide valuable insight for interpreting the physics of NRP discharges.

A small number of prior studies have investigated a limited range of characteristics of a single-pulse ns-discharge. For the single-pulse pin-to-pin ns-discharges, Lo[27] studied pin-to-pin electrode configuration with a 3 mm interelectrode gap energized by 25 ns and 54 mJ high voltage pulses. The electron number density up to 320 ns after the initiation of discharge and the electron temperature up to 40 ns were determined using Stark broadening, which yielded peak values of about $9.2 \cdot 10^{18}$ cm$^{-3}$ and 45,000 K, respectively. The measurements were taken at the core of the plasma filament with a spatial resolution of approximately 10 μm and temporal resolution of 5 ns. Additionally, gas temperature of up to 1,200 K at 15 ns after the initiation was measured using OES of the 2$^{nd}$ positive system of $N_2$. In another work by Miles[28], single pulses in a 2 mm interelectrode gap with 8 mJ pulse energy, and 55 ns pulse duration were studied. The resulting electron number density and electron temperature were measured by laser Thomson scattering and peaked at approximately $10^{17}$ cm$^{-3}$ and 3.2 eV, respectively, at 500 ns after the pulse initiation at the mid-point of the 2 mm gap with a spatial resolution of 150 μm.

In this work, a comprehensive, multi-parametric study of single-pulse ns-discharges in the pin-to-pin electrode configuration at atmospheric conditions was conducted for interelectrode gap sizes in the range 2-10 mm. Discharge voltage, current, rotational/vibrational temperatures, local gas density, and electron number density were determined and analyzed under different discharge conditions including gap distance between electrodes and level of energy deposition.

# Methodology and Experimental Setup

A detailed schematic of the experimental setup is shown in Figure 1. The electrodes, HV power supply, and voltage/current probes are shown in blue. The nanosecond HV pulse was produced by an Eagle Harbor NSP-3300-20-F pulser, where the peak voltage was set to be approximately 25 kV and the pulse width to about 90 ns. Two leads representing the positive and negative polarity outputs are each connected to tungsten pin electrodes with a tip radius of curvature of approximately 100 μm. Since the pulser generates a floating output signal, the actual voltage applied to the electrode assembly was measured by two HV probes where each of the probes was connected to the corresponding output line and shared same ground. The voltage probe model Tektronix P6015A (maximum rating of 30 kV and a bandwidth of 60 MHz) was utilized in this experiment. The current was recorded by a high-frequency current transformer (Bergoz FCT-028-0.5-WB) with a bandwidth of 1.2 GHz connected on the positive lead of the pulser near the electrodes. All signals were recorded by high bandwidth oscilloscopes (Lecroy HDO3904, 3 GHz bandwidth, 40 GS/s sampling rate and Lecroy WavePro 735Zi, 3.5 GHz bandwidth, 40 GS/s sampling rate). The HV pulses were sent to the electrodes at a maximum pulsing frequency of 1 Hz to ensure no memory effects from the previous discharge pulses. To improve the pulser impedance matching during the discharge, an additional matching resistor stage (a 5 kΩ resistor connected in parallel and two 260 Ω resistors connected in series) was utilized. The gap distance between the electrodes was adjusted in the range 2-10 mm and measured by a caliper. To adjust energy deposition into the discharge, additional resistors $R$ of 200 or 400 Ω were added in series with each discharge electrode as shown in Figure 1.

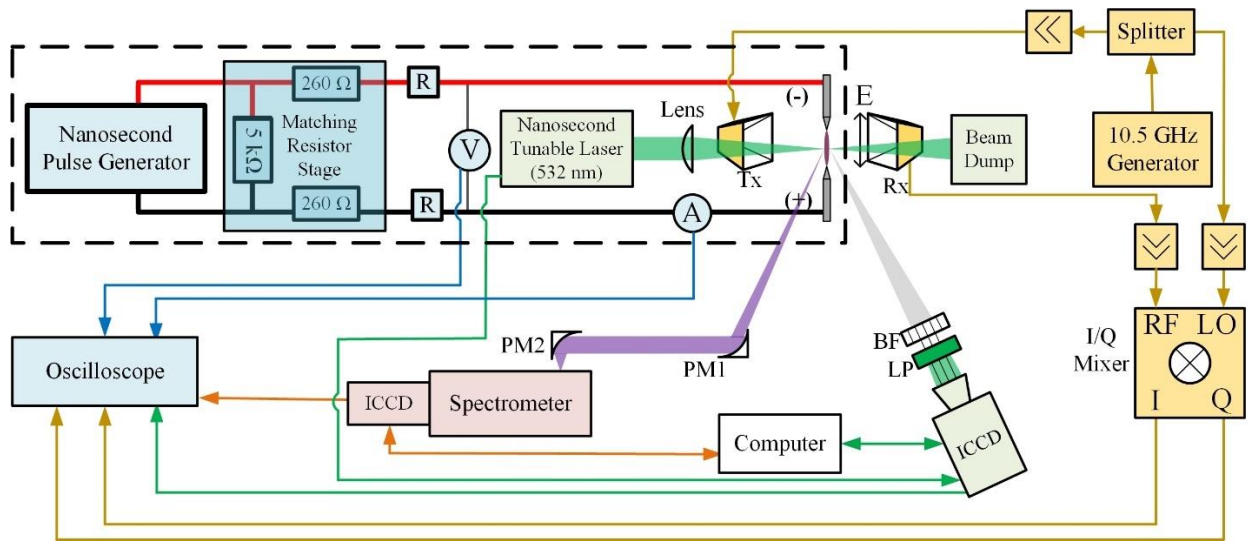

**Figure 1. Schematic of the experimental setup. (Blue) Pulse generator and voltage/current diagnostics. (Green) Laser Rayleigh scattering equipment. (Red/Orange) OES equipment. (Yellow) Microwave Rayleigh scattering equipment.**

The OES temperature measurement portion of the experimental setup is shown in Figure 1 and highlighted in red and orange. In this measurement, emission of the 2$^{nd}$ positive system (SPS) of nitrogen with $\Delta v = v' - v'' = -2$ (corresponding to the relaxation of electronically excited nitrogen molecules from the C to B state) was recorded and fitted in SPECAIR to evaluate the rotational, $T_{rot}$, and vibrational, $T_{vib}$, temperatures of the gas.[29,30] OES of the SPS of $N_2$ has been widely used previously for the determination of temperatures of air and $N_2$ plasmas.[27,31–34] $T_{gas}$ was taken to be equal to $T_{rot}$ in these measurements based on fast rotational–translational relaxation and the predominant creation of nitrogen $N_2$(C) by electron-impact excitation of $N_2$(X).[35] It has to be noted that the $T_{vib}$ measured and presented in this work is the vibrational temperature for $N_2$ in the C state rather than the X ground state. From the recorded spectrum, one can estimate $T_{vib}$(X) from $T_{vib}$(C) through the Franck-Condon coefficients, assuming that the majority of $N_2$(C) is created by electron-impact excitation of $N_2$(X).[36] The plasma emission at the mid-point of the gap between the electrodes was captured by a dual-parabolic-mirror system creating 1:1 image of the plasma channel on a 200 μm slit of a spectrometer (Princeton Instruments SP-2750i) and then recorded by an ICCD camera (Princeton Instruments PI-MAX4). For measurements during the ns-discharge, an ICCD gate width of 5 ns and a time step of 5 ns was used. For measurement of gas temperature after the discharge, OES enhanced with a probing ns-pulse was utilized.[35] The idea of the method is based on using the probing pulse to excite the emission of the SPS of $N_2$ at the desired moment of time while ensuring that heating by the probing pulse itself is negligible. For all OES spectra, 50-200 accumulations were taken to enhance signal-to-noise ratio. A detailed description of the method can be found elsewhere.[35]

For visualization of the plasma discharge, the ICCD camera was placed directly in front of the discharge to record the broadband light emission. The diameters of the plasma channels (*D*) were estimated from the recorded images.

For the evaluation of the local bulk gas density $n_g$, Laser Rayleigh scattering (LRS) was utilized. LRS is a widely-used technique for the determination of gas density with notable prominence in the fields of aerodynamics and combustion.[37–41] After applying a laser beam into a gaseous object of interest, the incoherent scattered signal of gas particles can be collected for the measurement of local gas density, as the signal intensity is proportional to the amount of gas particles within the probe volume. The intensity of the LRS signal is also a function of local gas temperature. However, the dependence between signal intensity and temperature is relatively weak (~1% per 1000 K).[42] As the bulk gas temperature measured in this study will be shown to be no more than 5000 K, it is assumed that the LRS signal is solely a linear function of local gas density. LRS portion of the experimental setup is depicted in Figure 1 and highlighted in green. A broadly-tunable Nd:YAG laser with Pellin-Broca spectral cleaning filter (EKSPLA NT342) was used to produce a 5 ns, 1±0.1 mJ laser pulse at 532±0.5 nm. The linearly-polarized beam was then focused in the center of the electrode gap perpendicular to the electrodes' axis using a lens with focal length of 175 mm, and the scattered signal was captured by an ICCD camera. Spatial resolution of the system was limited to the beam waist size of 320 μm. A 532±10 nm band pass filter and film polarizer was installed at an angle perpendicular to the laser beam/plasma filament for improved LRS signal isolation. Even though the LRS signal overlaps with Thomson and Raman scattering, the contribution of Rayleigh scattering was dominant for the experimental conditions in this work when $n_e < 10^{16}$ cm$^{-3}$ as described in Ref. [43]. Scattered laser signals were recorded for a range of times (100 ns to 1 ms) after the discharge. Accumulations of 10 events were taken to improve the signal-to-noise ratio and mitigate laser variance. The measurement of $n_g$ through LRS was further applied to Equation (2) below for the evaluation of collisional frequency $v$, and ultimately for the evaluation of $n_e$ by microwave Rayleigh scattering which will be discussed in detail in the following paragraphs.

For microwave Rayleigh scattering (MRS) measurements of total electron number $N_e$ and electron number density $n_e$, a homodyne microwave scattering system at 10.5 GHz was constructed. The output signal $U_{out}$ of the MRS signal was calculated from the outputs of the I/Q mixer utilized in the detection circuit: $U_{out} = \sqrt{\delta I^2 + \delta Q^2}$. The MRS system is shown in Figure 1 and highlighted in yellow. MRS is a non-intrusive diagnostic technique first proposed by M. N. Shneider and R. B. Miles for the measurement of electron number density of microplasmas at atmospheric conditions.[44] The technique is principally based on elastic microwave-plasma scattering in the quasi-Rayleigh regime referring to utilization of elongated plasma channels with diameter << microwave wavelength and collision-dominated electron motion. This regime is equivalent to classical Rayleigh scattering if incident microwave radiation is linearly polarized along the plasma channel. The scattered microwave intensity from the collisional-bound electrons can thus be a measure of the total number of electrons inside the plasma body as the scattered E-field is proportional to the electron count (coherent scattering). This method has been successfully applied to a variety of microplasmas including laser induced plasmas, helium plasma jets, and pin-to-pin NRP discharges.[26,34,45–47] The expressions for output signals of the MRS system are as follows:[44]

$$U_{out} = \begin{cases} A\frac{e^2}{mv}N_e = A\frac{e^2}{mv}n_e V & \text{— for plasma} \\ AV\varepsilon_0(\varepsilon - 1)\omega & \text{— for dielectric bullet scatterer} \end{cases} \quad (1)$$

where *e*: electron charge
*m*: electron mass
$\varepsilon_0$: dielectric permittivity of vacuum
$\varepsilon$: relative permittivity of the dielectric bullet
$\omega$: microwave frequency
$v$: collisional frequency
*A*: proportionality factor
*V*: volume of the plasma/dielectric scatterer
$U_{out}$: signal detected by the MRS system
$N_e$: total number of electrons
$n_e$: electron number density

For a ns-discharge in air with relatively low degree of ionization, the collisional frequency *v* can be calculated as:[48]

$$v[s^{-1}] = n_g \sigma v_{Te} = 1.95 \cdot 10^{-10} \cdot n_g[cm^{-3}] \cdot \sqrt{T_e[K]} \quad (2)$$

For the evaluation of collisional frequency, $T_e$ was assumed to decay from 10,000 K down to 2,000 K within the first 1 μs and then stay constant for the remainder of the plasma lifetime.[28,49] For the duration of the ns-pulse, $n_g$ can be assumed constant throughout the discharge. However, it has been demonstrated that the pin-to-pin nanosecond discharges considered in this study have energy depositions on the order of ~5 mJ and a corresponding gas temperature rise of multiple thousands of Kelvins is expected.[35] It is thus necessary to treat $n_g$ as a variable after the ns-discharge when calculating $n_e$, hence, the temporal measurement of gas density is required and was accomplished using LRS technique. More details about the necessity to account for $n_g$ variation from the unperturbed atmospheric level due to the ns-discharge and validation of combined MRS-LRS diagnostics can be found elsewhere.[43]

A calibration of the MRS system was conducted with a Teflon cylinder (1/8'' diameter, 1 cm length) to determine the coefficient *A* based on the bottom expression in the Equation (1). The calibration coefficient *A* was then applied to the ns-discharge plasma using top expression in Equation (1). The corresponding $U_{out}$ signal was utilized along with collisional frequency *v* derived from LRS measurement, and, finally, $N_e$ was calculated. With the estimated diameter of the plasma based on ICCD images, the spatially-averaged electron number density $n_e$ was calculated as: $n_e = \frac{N_e}{\frac{\pi}{4}D^2 d}$, where *d* is the gap distance between electrodes and *D* is the diameter of the plasma channel.

It should be noted that certain limitations apply for the MRS technique. Firstly, the plasma object must be placed far enough from the radiating horn antenna to ensure phase-free electric field radiation along the length of plasma. Secondly, the receiving horn must be placed at least 2$\lambda$ away from the plasma object to ensure far field radiation (~ 6 cm). In this study, both radiating and receiving horn antennas were placed 10 cm away from the plasma sample to meet the requirements above. Furthermore, the plasma diameter needs to be smaller than the skin layer depth to prevent shielding and ensure a uniform E-field distribution across the plasma channel. For the conditions of this experiment (plasma channel with ~300 μm diameter), the upper limit of the $n_e$ that a 10 GHz RMS system can measure is around $10^{16}$ cm$^{-3}$. Lastly, the elongated shape of the plasma channel eliminates depolarization effects and allows for considering the E-field inside the plasma channel to be equal to that in the incident wave.[45]

# Results and Discussion

## 1. Discharge Regimes

Two distinct discharge regimes were observed in this work: spark and corona. The experiments were conducted in ambient air at $p$=1 atm and transition between the spark and corona regime was observed by adjusting the gap distance between the electrodes $d$. Smaller discharge gaps ($d <$ 6 mm) were associated with the spark regime, while larger gaps ($d >$ 8 mm) were associated with the corona regime. Discharges with a gap distance of 7 mm were not repeatable, occurring either as spark or corona and, therefore, experimental data for $d$ = 7 mm was excluded from the analysis.

One can distinguish between the two regimes using several different discharge characteristics, including the discharge voltage ($V$), current ($I$), energy deposition, total electron production, and plasma images. Figure 2 shows an example of the *V-I* waveforms for corona, spark, and no-discharge cases. The *V-I* waveforms for the corona discharge are very similar to those where no discharge occurs. This is due to the fact that the electron number density of the plasma is relatively low in the corona regime, and thus the conductivity of the gas between the electrodes is fairly small. This results in a current ($<$ 2A) comparable to the displacement current flowing through the capacitor formed by the discharge electrodes without any discharge. Whereas for a spark discharge, a highly conductive plasma channel is established, resulting in a sudden drop of the voltage between electrodes and a very high level of discharge current ($>$ 10 A).

Visual observation of the discharge reveals significant differences between the corona and spark regimes. It was observed that the corona regime ($d >$ 8 mm) consists of multiple transient streamers that are constantly igniting, dying out, and randomly moving between the electrodes. The ICCD images during the first few nanoseconds after breakdown show a cathode-directed streamer initiated from the anode with the ionization front propagating at a speed of about 1 mm/ns. Note that for a corona discharge of stronger intensity (observed at smaller gaps ~8 mm), one may observe with a naked eye (or with long exposure time if photographed) a discharge appearance similar to that of the glow discharge.[19] Temporally-resolved photography shows that the plasma volume is still comprised of a large number temporally and spatially evolving streamers. In contrast, in the spark regime ($d <$ 6 mm), the ICCD image in Figure 2(c) shows that the plasma channel in spark regime was a single, thin, and highly-luminous (about 6 times more intense than corona) filament connecting the electrodes. The breakdown with $d < 6$ mm gaps was occurring almost instantly, bridging the entire gap with bright filament on the time scale faster than 3 ns (maximal temporal resolution of the ICCD camera used in this work). This suggests involvement of a very fast breakdown mechanism (substantially faster than a streamer breakdown observed at larger gaps $d > 8$ mm). Note that, for parallel-plate geometry, the range of $p·d < 200$ Torr·cm is associated with the Townsend breakdown mechanism of avalanche multiplication, while $p·d > 4000$ Torr·cm corresponds to the streamer breakdown mechanism.[48] In this work the streamer breakdown was observed for the gap distance above ~ 7 mm ($p·d \geq 500$ Torr·cm), while smaller $p·d$ values were associated with an even faster breakdown mechanism than streamer or Townsend breakdown. The mechanisms of such extremely fast breakdown events are described in more details in Refs. [48,50,51]. Table 1 below summarizes the main discharge properties of the two regimes observed in this work. More details on each discharge regime are given in the following sections.

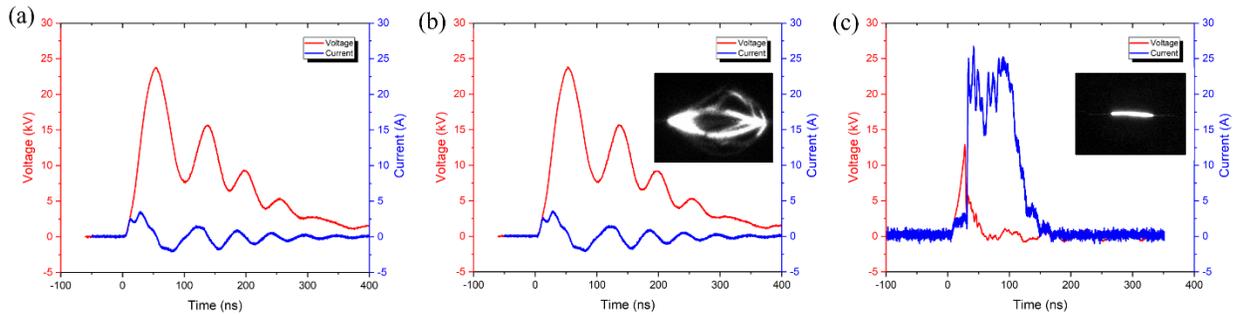

**Figure 2. V-I waveform of a single ns-pulse: (a) without breakdown (with dielectric material placed between the electrodes); (b) in the corona regime ($d$=10 mm); (c) in the spark regime ($d$=2 mm). ICCD images for corona and spark regimes are also shown to the right of the V-I curves for each regime.**

**Table 1. Summary of the discharge properties in the two regimes observed in this work: spark ($d \leq 6$ mm) and corona ($d \geq 8$ mm).**

| Regime | Gap distance $d$, mm | Breakdown voltage $V_{br}$, kV | Peak current $I_{peak}$, A | Energy deposition, mJ | Total electron production, $(N_e)_{max}$ | Gas temperature $T_{gas}$, K | Light emission |
|---|---|---|---|---|---|---|---|
| Spark | 2 - 6 | < 22 | > 10 | > 2 | > $1\times10^{12}$ | > 1000 | Single thin plasma channel; intense light emission |
| Corona | ≥ 8 | ~ 25 | < 2 | < 1.5 | < $3\times10^{11}$ | < 360 | Multiple streamers; weak light emission (6x weaker than spark) |

## 2. Spark Regime

## 2.1 Typical discharge parameters

Figure 3 is an example of the temporal evolution of voltage $V$, current $I$, $n_e$, $T_{rot}$, and $T_{vib}$ for a spark discharge with $d = 5$ mm. One can see that spark breakdown occurred at a voltage of around 19 kV and $t \approx 26$ ns as indicated by a steep drop of the interelectrode voltage to the 5-10 kV range, peak current of approximately 17 A, and creation of electrons in the gap with average number density of about $n_e=1.5\times10^{15}$ cm$^{-3}$. Additionally, one can observe a second breakdown event (BD2) occurring at $t\approx70$ ns. It is indicated by a further steep drop of the voltage to less than 1000 V, second current peak of about 17 A, and additional increase of $n_e$ to about $5\times10^{15}$ cm$^{-3}$. BD2 is likely triggered by the increase of the electric field in the cathode vicinity which is expected due to constant removal of electrons from the plasma channel to the anode to support the discharge current conduction and the corresponding buildup of positive ion charge on the cathode-facing side of the plasma column. Thus, BD2 is potentially associated with cathode sheath breakdown, ignition of a cathode spot, and transition of the discharge to cathodic arc regime.[48,52] This stage of the discharge is expected to be destructive for the electrode assembly due to erosion of cathode material from the cathode spots.

Figure 3 also shows the temporal evolution of $T_{rot}(=T_{gas})$ and $T_{vib}$. One can see that $T_{gas}$ increased from room temperature to around 1750 K at 80 ns. $T_{vib}$ increased from 4,000 K to 10,000 K within the same time frame. The temperatures cannot be measured by the OES technique beyond $t = 80$ ns due to the lack of SPS emission at these times. This can be explained by reduction of the local reduced field $E/n_g$, where $E$ is the electric field strength and $n_g$ is the gas number density. Specifically, reduced fields $E/n_g >10^{-15}$ V·cm$^{-2}$ are required for efficient pumping of electronically and vibrationally excited states of $O_2$ and $N_2$, whereas efficiency of these transitions decreases for lower reduced fields.[48] A quick estimation of the reduced field in current experiment indicates that $E/n_g$ drops below $10^{-15}$ V·cm$^{-2}$ (neglecting voltage drop in the cathode sheath) at 20 ns after the discharge initiation.

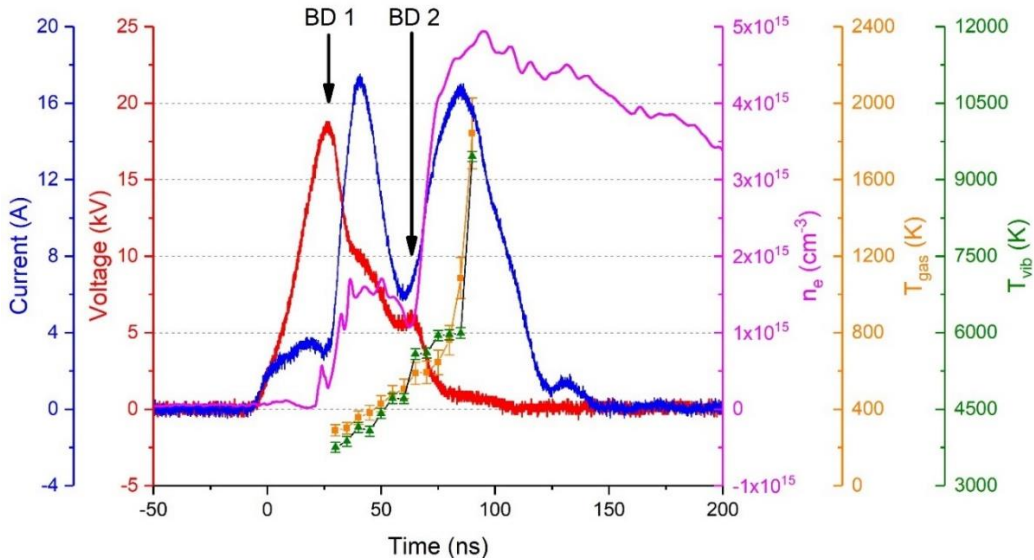

**Figure 3. Temporal measurement of discharge parameters within the pulse at $d = 5$ mm. Moments of two breakdowns (BD) events are also labeled.**

The temporal evolution of the gas temperature after the ns-discharge was measured by implementing the probing pulse method[35] and is presented in Figure 4. Note that the earliest measurement of $T_{gas}$ was conducted at $t=10$ μs which is limited by the highest repetition frequency of 100 kHz that the pulser can produce. The gas temperature measured at $t=10$ μs is significantly higher than the maximum temperature during the pulse (3500 K vs. 1750 K as shown in Figure 3), indicating that the gas underwent further heating after $t>100$ns. This additional heating is mainly due to the vibrational relaxation of excited species (mainly $N_2$) which lasts for several microseconds.[3] $T_{gas}$ reduces the room temperature by about 1 ms after the ns-discharge.

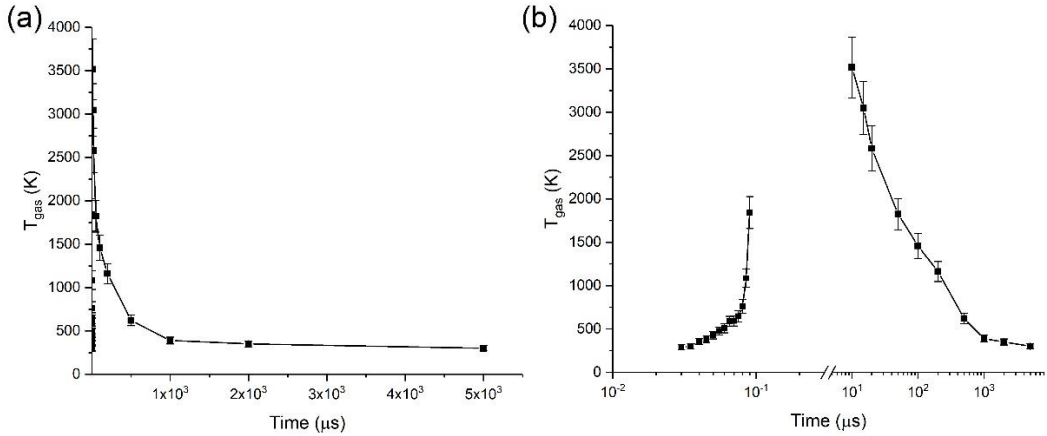

**Figure 4. Temporal evolution of gas temperature $T_{gas}$ within the pulse and post-discharge. (a) linear scale; (b) log scale**

Temporal evolution of relative gas density $n_g/n_{g0}$ measured at the center of electrode gap (where $n_{g0} \approx 2.5 \times 10^{19}$ cm$^{-3}$ is the gas density at standard atmospheric conditions) is plotted in Figure 5(a). Figure 5(b) and (c) show $n_g/n_{g0}$ profiles measured by LRS in the radial direction at the midpoint between electrodes (1-5 μs, and up to 1 ms post-discharge, respectively). It can be seen that $n_g$ drops to 30% of the ambient level within the initial 1 μs post-discharge, stays constant for about 50 μs, and recovers to ambient level in the next 1 ms. Additionally, one can observe a pressure wave propagating outward at a speed of approximately 500 m/s in Figure 5(b).

Interpretation of the measured scattered signal may be affected by dissociation of oxygen.[53–56] Considering an extreme case of complete oxygen dissociation, one may expect LRS measurement error of about 15%,[43] and, in this work, we have accounted for the possibility of oxygen dissociation via corresponding error bars. One should expect a substantially lower dissociation degree under the conditions of our experiments given that only 30% dissociation was reported for ns-discharges at significantly larger discharge pulse energy of 20 mJ.[57]

The temporal evolution of the plasma diameter/volume was estimated visually from line-of-sight ICCD images, and the results are shown in Figure 6. The diameter of the plasma channel increases from about 250 μm up to 0.9 mm within the first 10 μs after the pulse and then stays approximately constant until the plasma fully decays. Curves smoothly connecting measured data points for $n_g$ and plasma diameter were used as input for evaluating $N_e$ and $n_e$ using Equation (1). Abel inversion was also performed for multiple raw ICCD images, and the estimated plasma diameter using the inverted images was within 5% of the diameter estimated from the raw images. Thus, plasma diameter was evaluated from raw line-of-sight ICCD images in this study and potential error was accounted via the uncertainty bars shown in Figure 6(a).

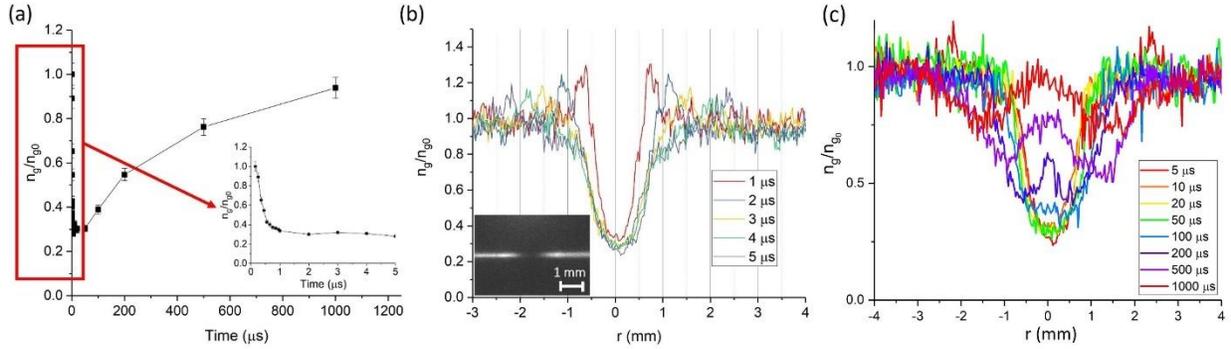

Figure 5. (a) Temporal evolution of $n_g$ measured by LRS. Radial profile of the relative density perpendicular to the plasma filament for (b) t = 1-5 μs after discharge, (c) 5-1000 μs after discharge. Reproduced from Ref. [43], with the permission of AIP Publishing.

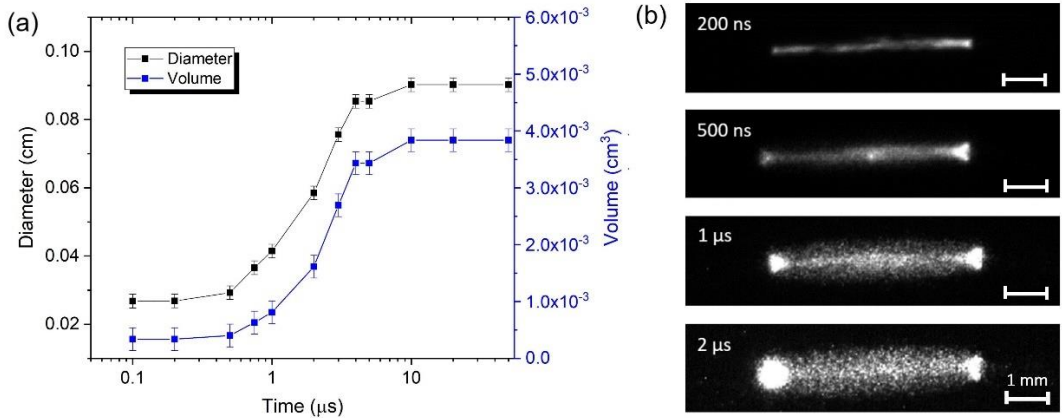

Figure 6. (a) Temporal evolution of plasma diameter and volume. Black: diameter; blue: volume. (b) ICCD images of the plasma after the ns-discharge pulse. Reproduced from Ref. [43], with the permission of AIP Publishing.

The temporal evolution of $N_e$ and $n_e$ is presented in Figure 7. The results are calculated using Equation (1) with the input data of $U_{out}$, $v$, and $V$ obtained using the procedure described in Ref. [43]. The collisional frequency $v$ was determined via local gas density $n_g$ measured by LRS. As one can see in Figure 7, the characteristic decay time of $n_e$ was about 150 ns. This corresponds to a decay rate of approximately $0.67 \times 10^7$ s$^{-1}$ which is about order of magnitude smaller than that reported for atmospheric pressure conditions in an earlier study ($0.76 \times 10^8$ s$^{-1}$).[48,58]

To understand this discrepancy, we will analyze the decay using a simple model considering only dissociative recombination and three-body attachment to oxygen reactions: $\frac{\partial n_e}{\partial t} = -\beta n_e^2 - \nu n_e$. This equation was modeled numerically using the measured $n_g$ evolution, dissociative recombination of oxygen ions ($e + O_2^+ \rightarrow O + O$) and three-body attachment to oxygen ($e + O_2 + O_2 \rightarrow O_2^- + O_2$ and $e + O_2 + N_2 \rightarrow O_2^- + N_2$) reactions with the following rate coefficients: $\beta = 2 \times 10^{-7} \times \frac{300}{T_e}$ [cm$^3$s$^{-1}$], $k_1 = 1.4 \times 10^{-29} \times \frac{300}{T_e} \times \exp\left(-\frac{600}{T_{gas}}\right) \times \exp\left(\frac{700 \times (T_e - T_{gas})}{T_e \times T_{gas}}\right)$ [cm$^6$s$^{-1}$], $k_2 = 1.07 \times 10^{-31} \times \left(\frac{300}{T_e}\right)^2 \times \exp\left(-\frac{70}{T_{gas}}\right) \times \exp\left(\frac{1500 \times (T_e - T_{gas})}{T_e \times T_{gas}}\right)$ [cm$^6$s$^{-1}$], and $\nu = k_1 n_{O_2}^2 + k_2 n_{O_2} n_{N_2}$.[59] A very important parameter of this modeling is electron temperature ($T_e$). The electron temperature is expected to be close to vibrational temperature of nitrogen in the afterglow phase, suggesting $T_e \sim$ several thousand K during first few microseconds right after the discharge pulse.[57] At the same time, a reasonable agreement between the model and measurements for the initial portion of the electron decay ($t <$ 500 ns) requires assuming higher values of $T_e$ (gas temperature in the range $T_{gas}$ =2000-4000 K was utilized based on the measurements). Specifically, $T_e$ =1 eV and 3 eV yield decay times of 30 ns and 80 ns, respectively. The observed discrepancy in the electron decay time can be potentially attributed to inaccuracies in rate coefficient approximations (note, the above approximations for the reaction rate coefficients were proposed for cold vibrationally unexcited gas $T_{gas} <$ 500 K)[59] and inability to spatially resolve the decay of the dense spark core by the MRS technique. (It is also worth noting, that recent measurements conducted using laser Thomson scattering technique indicate anomalously slow $T_e$ decay from ~ 3 eV right after the discharge pulse to ~ 1 eV at 1μs after the discharge pulse, for 55 ns FWHM and 8 mJ discharge pulse.[28]) Additionally, the obtained results indicate that the initial stage of the decay (up to $t \sim$ 1-2 μs) is likely to be governed by comparable contributions of both dissociative recombination and attachment to oxygen mechanisms, while the later stage of the decay is expected to be dominated by the attachment to oxygen.

To summarize, the reduction of local gas number density at the spark location right after the ns-discharge pulse is an important factor contributing to the anomalously slow electron decay observed in the experiments (~150 ns). Precise knowledge of electron temperature dynamics along with exact reaction rate coefficients are also required for correct decay interpretation.

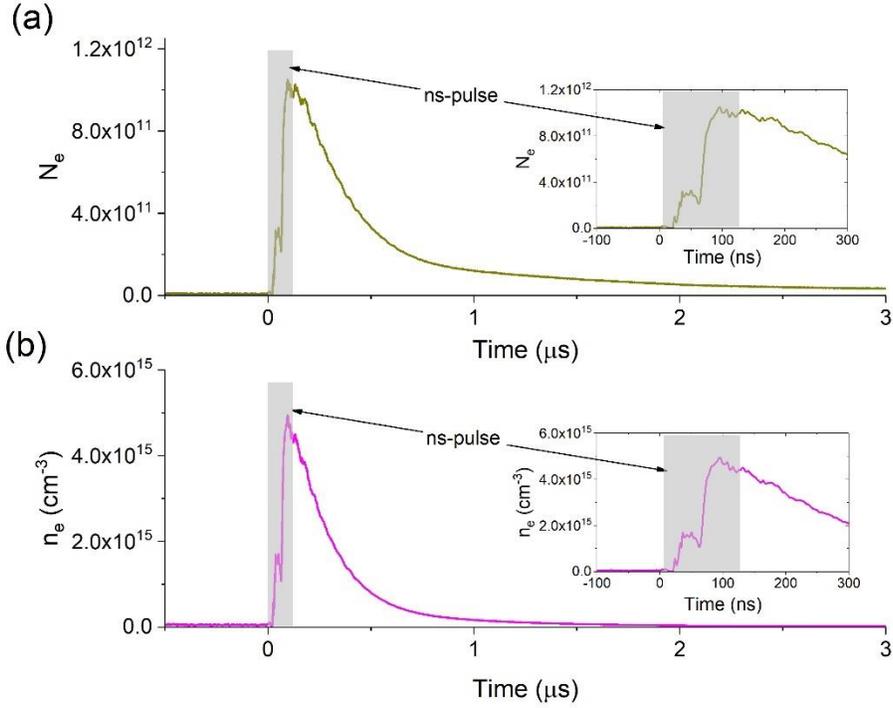

**Figure 7.** Temporal evolution of $N_e$ (top) and $n_e$ (bottom). The insert images show the first 300 ns of the discharge in greater detail. The grey area represents the duration of the ns-pulse.

## 2.2 Effect of pulse energy on discharge parameters

Next, we investigate how the ns-pulse energy affects the discharge parameters in the spark regime ($d$=5 mm). In this study, the amount of energy deposition to the discharge was controlled by adding current-limiting resistors in series with each discharge electrode. The energy per pulse was calculated by integrating the product of voltage measured across the interelectrode gap and current over the entire pulse time. By adjusting the current-limiting resistors between $R_{CL}$=0, 200, and 400 Ohm, the total discharge pulse energy was measured to be $\mathcal{E}_{tot}$=5.04 (±0.14), 3.76(±0.10), and 3.07 (±0.08) mJ, respectively. Temporal evolution of discharge parameters including voltage, reduced electric field (neglecting voltage in cathode sheath), current, discharge energy (from the beginning of the ns-pulse until current moment of time), temperatures (gas/rotational, vibrational), $N_e$, and $n_e$ are presented in Figure 8 for the duration of the ns-pulse.

Figure 8 (a) shows that the initial breakdown voltage is approximately the same for all three discharge energies at around 18 kV. This is due to the fact that prior to breakdown the capacitive impedance of the air gap between the electrodes is substantially larger ($\sim \tau/C$=2.5 kΩ, where $\tau \approx$10 ns is the characteristic time of the voltage rise and $C \approx$4 pF is the capacitance of the electrodes) than the resistance of the current-limiting resistors used. Thus, before breakdown, the impedance seen by the pulser does not change significantly by introducing different current-limiting resistors. Therefore, similar voltage rise times (within 10%) are produced by the pulser regardless of the $R_{CL}$ value and the voltage is applied almost entirely to the interelectrode gap. Correspondingly, energy stored in the capacitor formed by the electrode assembly prior to the breakdown (until $t$~26-30 ns) is very similar for all three $R_{CL}$ values used, ~0.64±0.03 mJ, as seen in Figure 8 (d).

Unlike the breakdown voltage, the shape of the voltage pulse differs more substantially after breakdown (at $t$~26-30 ns) for the different discharge pulse energies used. This is caused by the more comparable range of impedances of the plasma to the current-limiting resistors. Indeed, after the breakdown, a highly-conductive/low-resistance plasma channel is formed, bridging the gap between the electrodes (~600-1000 Ohm after the initial breakdown and then dropping to <120 Ohm after the second breakdown based on the waveforms shown in Figure 8(a) and (c)). Clearly, comparable range of utilized $R_{CL}$ to that of the plasma channel is achieved after the breakdown. The comparable impedances of the plasma and the resistors makes adjustment of $R_{CL}$ an efficient way to redistribute the ns-pulse energy between the discharge and the current-limiting resistors. Specifically, utilization of larger $R_{CL}$ causes a larger fraction of the energy to be deposited in the current-limiting resistor and less in the discharge.

One can also observe in Figure 8(a), (b) and (c) a substantially delayed BD2 time for lower pulse energies while the interelectrode voltage just prior to BD2 was very similar (~6 kV) regardless of the discharge pulse energy. Specifically, BD2 occurred at 65 ns, 75 ns, and 82 ns for the pulse energies 3.07, 3.76 and 5.04 mJ, respectively. This can be explained by larger discharge currents generated at higher pulse energies and, therefore, faster removal of electron charge from the plasma column. This in turn causes a corresponding faster build-up of ion charge on the cathode-facing side of the plasma column and growth of electric field to a level sufficient for breakdown of the cathode sheath. Figure 8(e) and (f) show the temporal evolution of $T_{gas}$ (= $T_{rot}$) during the pulse. It can be seen that gas heating is more rapid and more substantial with increasing discharge pulse energy. Indeed, $T_{gas}$ reached 1000 K in 70 ns, 95 ns and 100 ns for the pulse energies 5.04 mJ, 3.76 mJ and 3.07 mJ, respectively. The maximum temperature measured during the pulse is also higher for higher discharge pulse energy. Note that the time window of detectable SPS emission was longer for lower discharge pulse energy (since high $E/n_g$ was preserved for longer time at lower discharge pulse energy), which allowed extending $T_{gas}$ measurements to longer times.

Temporal evolutions of $N_e$ and $n_e$ during the ns-pulse are presented in Figure 8 (g) and (h). One can see that the number of electrons created in the gap after the initial breakdown BD1 (at $t$~26-30 ns) increased with discharge pulse energy. Specifically, deposition of approximately 1.5 mJ after BD1 (observed for the case with $\mathcal{E}_{tot}$ =3.07 and 3.76 mJ) led to establishment of quasi-steady-state electron number $N_e$=1.9×10$^{11}$ and electron number density $n_e$=1×10$^{15}$ cm$^{-3}$ (at $t$ ~ 40-60 ns) while energy deposition of ~2 mJ ($\mathcal{E}_{tot}$=5.04 mJ) yielded $N_e$=3.0×10$^{11}$ and $n_e$=1.6×10$^{15}$ cm$^{-3}$. This quasi-steady-state period of the discharge ($t$ ~ 40-60 ns) is likely associated with a balance between the electron production in the discharge volume ($\alpha V_{dr} N_e$, where $\alpha$ is the ionization coefficient and $V_{dr}$ is the electron drift velocity) and electron removal to the anode ensuring conduction of the discharge current ($I/|e|$; $I$ ~10A at $t$~50 ns). This balance yields $\alpha/n_g$~10$^{-18}$-10$^{-}$ [19] cm² and reduced electric field on the order of $E/n_g$~100-150 Td which is fairly consistent with the estimation based on measured discharge voltage.[60] Later breakdown of the cathode sheath (BD2) results in a significant (about 5-10 times) drop in the voltage (from about 6 kV before BD2 to about 500-1000 V after BD2). The corresponding substantial decrease of $E/n_g$ nearly stops the electron production inside the plasma volume due to the sharp reduction of the ionization coefficient with the decrease in $E/n_g$ (e.g., ionization coefficient drops 100 times when $E/n_g$ reduces from 100 to 60 Td).[60] Instead, electron number and electron number density in the gap are governed by production of electrons in the cathode spot vicinity, their transport through the gap and collection at the anode. This stage of discharge largely resembles a conventional cathodic arc and would be converted into a DC cathodic arc if discharge-driving voltage is maintained at the electrodes.

For all the discharge pulse energies considered, the majority of the energy was deposited to the interelectrode gas near and right after the time of initial breakdown BD1, while only incremental further energy deposition is associated with the second breakdown (and transition to cathodic arc). Specifically, 85-89% of the pulse energy was deposited into the gas before BD2, while only 11-15% was deposited after BD2. This shows that the fast ns-discharge is a substantially more efficient way to deposit energy into the interelectrode gas in comparison with a cathodic arc.

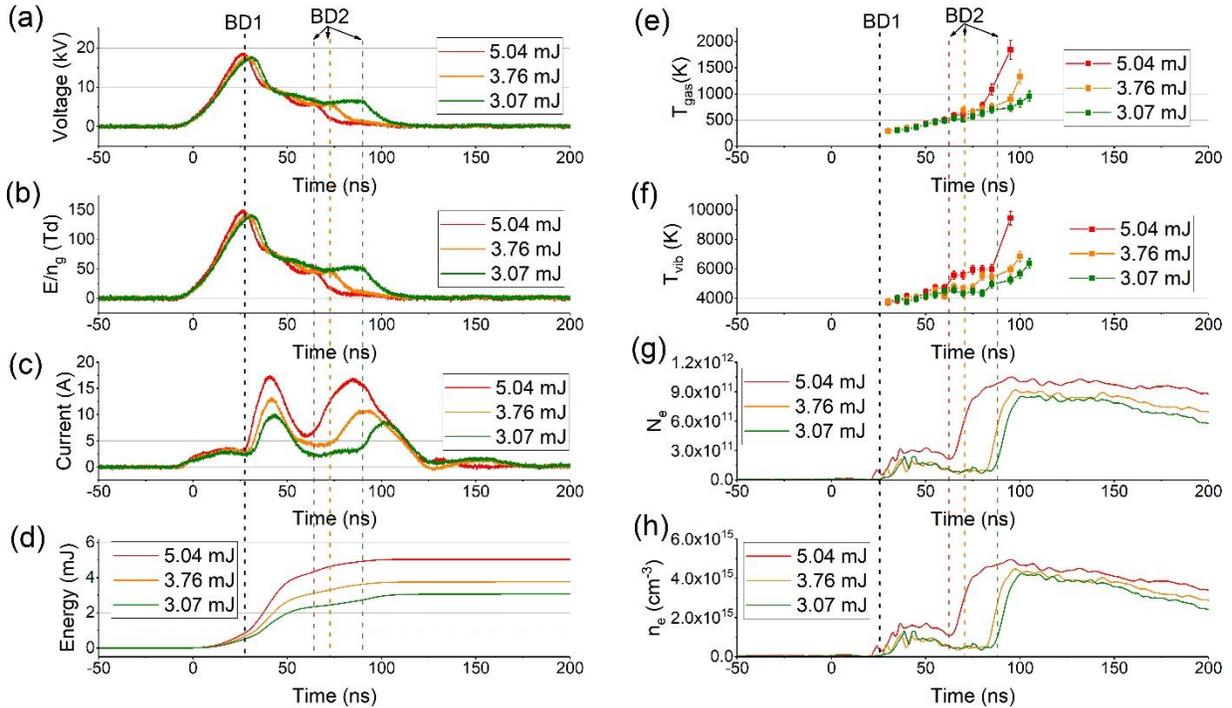

**Figure 8. Temporal measurement for different energy deposition. (a) voltage; (b) reduced field; (c) current; (d) energy; (e) gas temperature; (f) vibrational temperature; (g) total electron number; (h) electron number density**

The complete temporal evolution of the gas temperatures, measured by OES and OES enhanced with a probing pulse, for three discharge energy levels is presented in Figure 9. At $t = 10$ μs, a peak temperature of 3500 K was measured for 5.04 mJ pulse energy in comparison with 3000 K for 3.07 mJ. In addition, $T_{gas}$ after the ns-pulse decreased at a faster rate for higher pulse energies until about 100 μs, when the temperature decay curves collapsed to for a single curve for all discharge pulse energies. The gas temperature recovered back to room temperature by about 1 ms for all three discharge energies.[35] This faster initial temperature decay for higher $\mathscr{E}_{tot}$ is potentially due to generation of stronger vortex rings and thus a more intensive air mixing/cooling process when higher energy is deposited in the gas.[61–63]

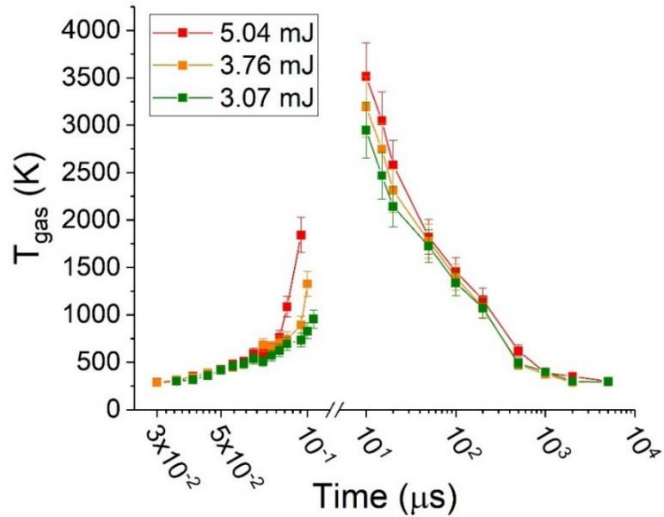

**Figure 9. Temporal evolution of $T_{gas}$ for different energy deposition during and after the HV pulse.**

The temporal evolution of the relative gas density ($n_g/n_{g0}$) during and after the HV pulse measured by LRS for different discharge pulse energies is presented in Figure 10(a). A more detailed plot of the initial dynamics of $n_g/n_{g0}$ (for $t < 5$ μs) along with the temporal evolution of the plasma diameter estimated from ICCD images is shown in Figure 10(b). Similar behavior of $n_g/n_{g0}$ for all three discharge pulse energies was observed, starting with a fast initial drop within 1 μs post-discharge. The relative gas density then remained nearly constant for 10s of μs followed by recovery to the ambient level in the next 1 ms. More significant reduction of gas density was found for higher discharge pulse energy; specifically, air density dropped to 55%, 50%, and 30% of ambient air density level for discharge pulse energies of 3.07 mJ, 3.76 mJ, and 5.04 mJ, respectively. It can be also seen that the plasma channels had very similar diameters regardless of the pulse energy level during and immediately after the pulse ($t < 500$ ns), but then expanded to larger diameters over the next 5 μs for higher pulse energy (650 μm for 3.07 mJ and 900 μm for 5.04 mJ).

Finally, the complete temporal evolution of $N_e$ and $n_e$ is plotted in Figure 11. It can be observed that the temporal decay of $n_e$ is elongated with increase of the pulse energy. This is due to achieving higher $T_e$ and a more substantial $n_g$ drop with higher pulse energy as described in Section 2.1.

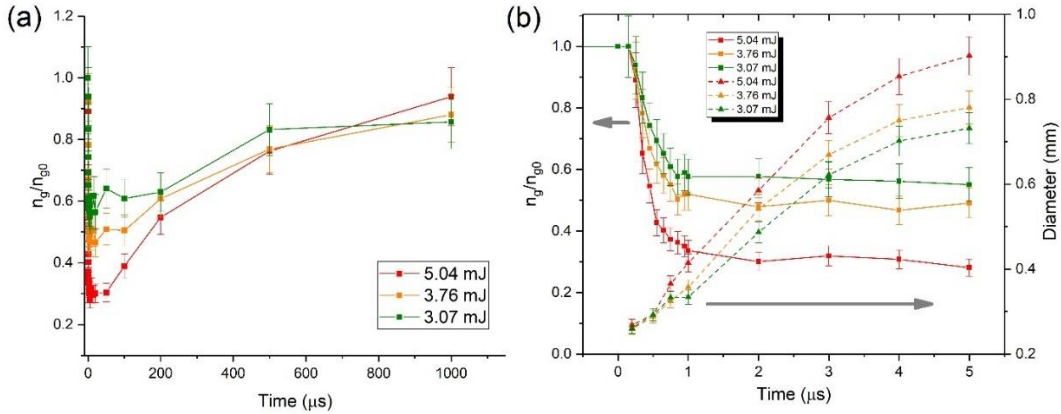

**Figure 10. (a) Temporal evolution of relative gas density for different energy deposition during and after the HV pulse. (b)Temporal evolution of gas density and plasma diameter for different energy deposition within the first 5 μs post-discharge. Solid line + square: relative gas density; dash line + triangle: plasma diameter.**

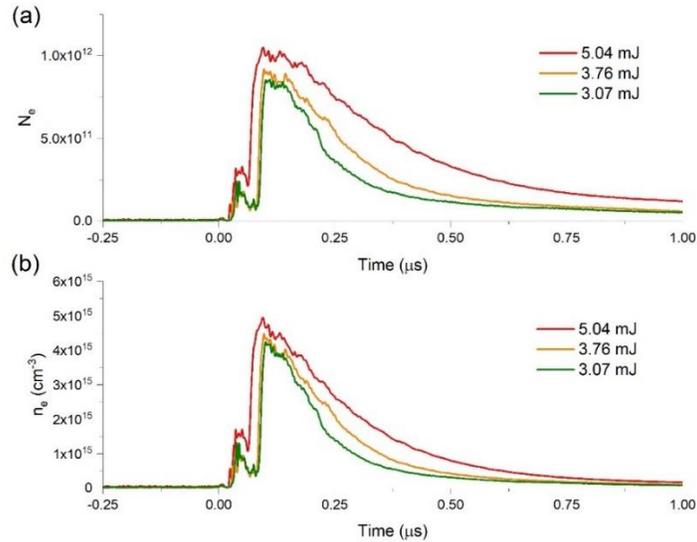

**Figure 11. Temporal evolution of (a) $N_e$ and (b) $n_e$ for different energy deposition during and after the HV pulse.**

## 2.3 Effect of gap distance on discharge parameters

Finally, we investigate the effect of the interelectrode gap distance $d$ on the discharge properties in the spark regime. (It is worth noting that the effect of gap distance should be ideally investigated when parameter of $\mathcal{E}_{tot}/d$ is kept constant. Even though it is not feasible to precisely preset the discharge pulse energy, the results of the experiments shown below will demonstrate that parameter $\mathcal{E}_{tot}/d$ was nearly constant.) Temporal evolution of the discharge properties for $d = $ 2, 4, 6 mm are presented in Figure 12 including voltage, reduced electric field $E/n_g$ (neglecting voltage in cathode sheath), current, discharge pulse energy per unit gap $\mathcal{E}/d$, $T_{gas}$, $T_{vib}$, $N_e$, and $n_e$. As one can see from Figure 12(a) and (b), the initial breakdown (BD1) of smaller gaps occurred at a lower voltage but higher reduced electric field. Indeed, as shown in Figure 12(b), $E/n_g$ at the moment of breakdown was 140, 160 and 260 Td for $d = $6, 4 and 2 mm, respectively. Additionally, the breakdown in smaller gaps occurred more abruptly (as indicated by more rapid voltage drop rate), generating higher discharge current peaks as shown in Figure 12(c). This correspondingly led to faster energy deposition per unit gap distance ($\mathcal{E}/d$) as shown in Figure 12(d) and caused faster gas heating which is confirmed by the $T_{gas}$ measurements in Figure 12(e). Higher discharge currents in smaller gaps led to earlier second breakdown BD2 associated with ignition of the cathodic spot, which is consistent with the results reported in the previous section. The total energy input per gap $\mathcal{E}_{tot}/d$ was rather consistent (~1 mJ/mm) for all gap sizes. Larger $E/n_g$ observed in the smaller gaps is expected to shift the balance between electron production in the discharge volume and electron removal to the anode to support current conduction toward production of higher electron number density and higher discharge current, and this is consistent with the experimental findings. The effect of gap distance on $V_{br}$, $(E/n_g)_{br}$, $I_{peak}$, $\mathcal{E}_{tot}$, $\mathcal{E}_{tot}/d$, $n_e$, $N_e$ (quasi-steady-state values after BD1) is summarized in Figure 13.

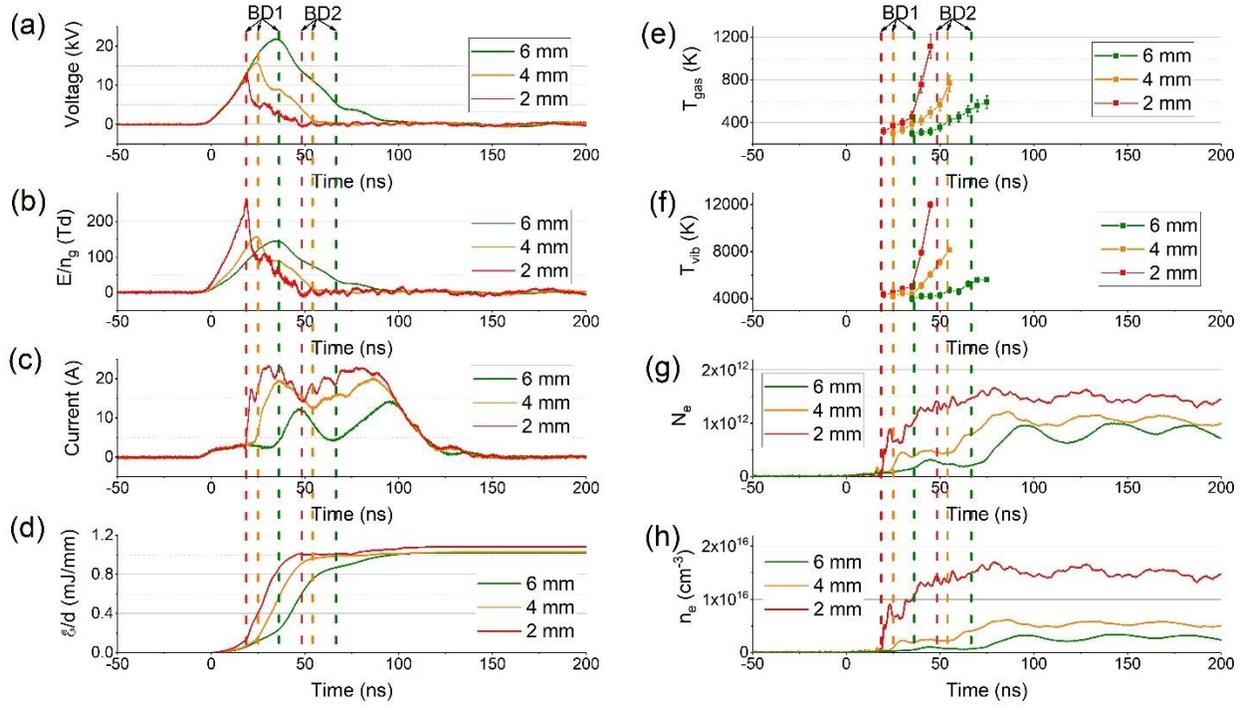

**Figure 12.** Temporal evolution of (a) voltage; (b) reduced field; (c) current; (d) energy; (e) gas temperature; (f) vibrational temperature; (g) total electron number; (h) electron number density for $d$ = 2, 4, 6 mm

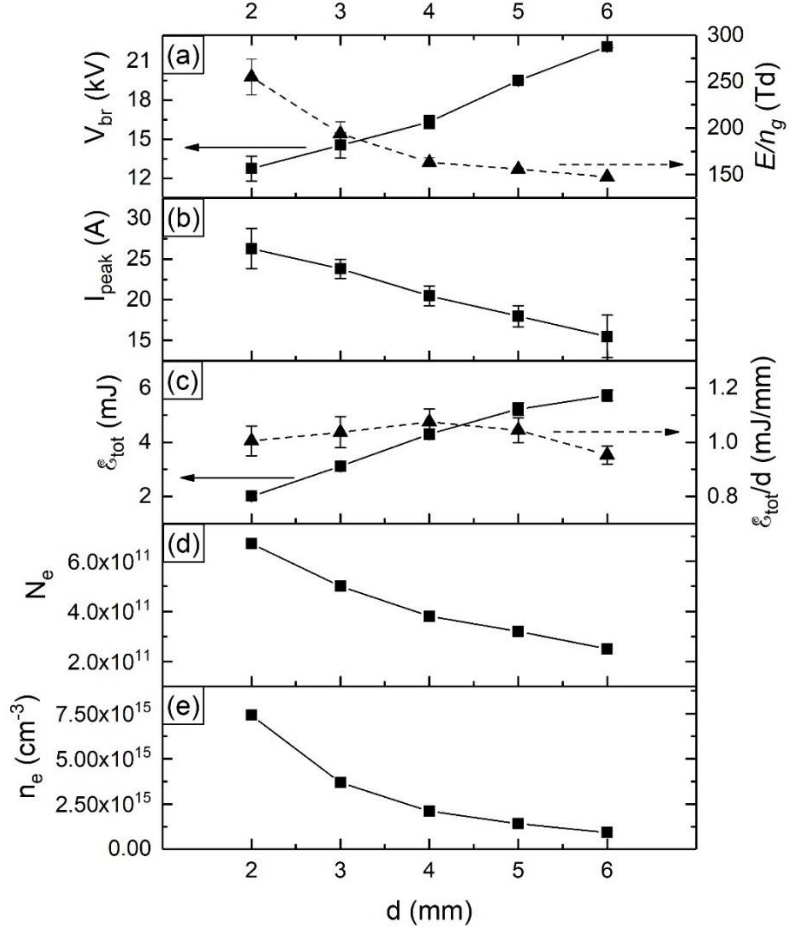

**Figure 13. (a) Voltage and reduced electric field at breakdown, (b) Peak current, (c) Energy input and energy input per unit gap length, (d) $N_e$ after BD1, and (e) $n_e$ after BD1 vs. interelectrode gap distance.**

The complete temporal evolution of $T_{gas}$ for $d$ = 2, 4, 6 mm are presented in Figure 14. One can see that the cooling rate of $T_{gas}$ at smaller gaps was faster than for the larger gaps. This can be potentially explained by a faster removal of hot gas between the electrodes when stronger vortex rings are formed in smaller gaps.[61–63]

Summarizing the measurements of plasma parameters after a ns-discharge pulse for tested pulse energies (0.6-1 mJ/mm) and gap sizes (2-6 mm), we conclude that complete recovery of the parameters to their unperturbed pre-discharge values occurs at ~ 1 ms after the discharge pulse for all cases (see Figure 4, Figure 5, Figure 9, Figure 10 and Figure 14). One can thus expect memory effects to occur at a pulse repetition frequency >1 kHz. Note that the 'cut-off' frequency indicating onset of memory effects and transition from the single-pulse ns-discharge regime to the NRP operational regime depends on many factors, including ambient temperature, pressure, pulse energy and shape, gas composition, and electrode geometry. Investigation of the NRP regime using the multi-diagnostics system developed in this work will be subject of future studies.

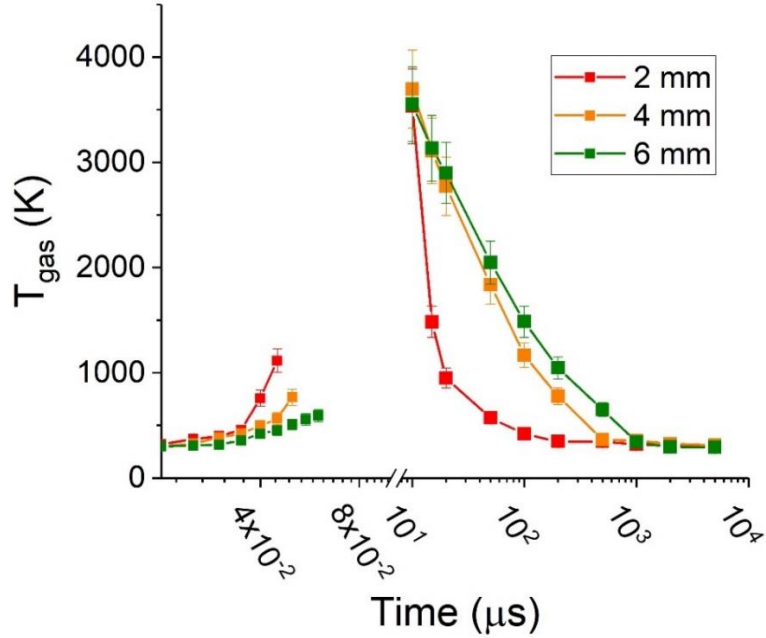

**Figure 14.** Temporal evolution of (a) $T_{gas}$ and (b) $T_{vib}$ during and after the HV pulse for $d$ = 2, 4, 6 mm.

## 3. Corona Regime and Comparison with Spark

The corona regime was investigated for gap distances of 8, 9, and 10mm. Voltage and current waveforms for $d$ = 8 mm are shown in Figure 15. In comparison to the *V-I* waveforms for the spark regime (e.g. Figure 3), it is noticeable that corona ignition was not associated with substantial voltage drops. On the contrary, the voltage waveform was similar to those observed when no breakdown occurs (see Figure 2), with a longer pulse duration than that set on the pulser (90 ns) and with significant voltage oscillations. This indicates that corona does not substantially affect the impedance of the interelectrode gap and thus the pulser is operating the entire time in the mismatched conditions being loaded with 5 kOhm active resistor of the matching resistor stage shown in Figure 1. The total measured current ($I$) consists of two components - displacement and conduction currents. Displacement current can be calculated from $I_{disp} = C \cdot \frac{dV}{dt}$ using the known capacitance of interelectrode assembly $C$=3.9 pF (determined from the experiments without breakdown). Finally, the conduction current can be determined as $I_{cond} = I - I_{disp}$ and the results are shown in Figure 15(b). One can see that the corona discharge current reached a peak value of approximately 1.7 A at $t$~50-70 ns.

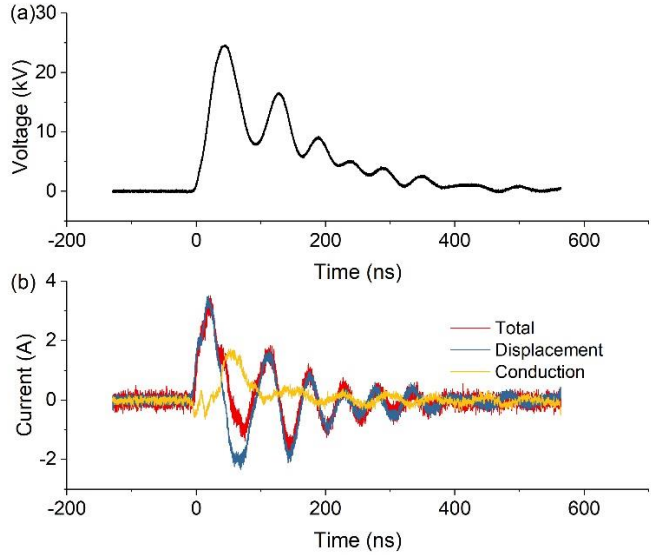

**Figure 15. Temporal evolution of (a) voltage and (b) current for corona regime at d = 8 mm**

The peak numbers of electrons ($N_{e,max}$) for $d$=8-10 mm are plotted in Figure 16 along with the corresponding peak values of $I_{cond}$. One can see that as $d$ increases, the peak conduction current decreases while the number of electrons produced increases. Both trends can be explained by the longer associated streamer length for larger $d$ and the correspondingly larger resistance and number of electrons. Note that due to the generation of multiple streamers between the electrodes and the random path of each individual streamer, confirmed using ICCD photography as shown in Figure 2(b), it was impossible to quantify the exact volume of the plasma channel, and, therefore, the value of $n_e$ was not calculated. The values of $T_{gas}$ and $T_{vib}$ temporally averaged over the discharge pulse duration were also measured using OES, and the results were 360 ± 20 K and 3700 ± 100 K, respectively, for all three gap distances.

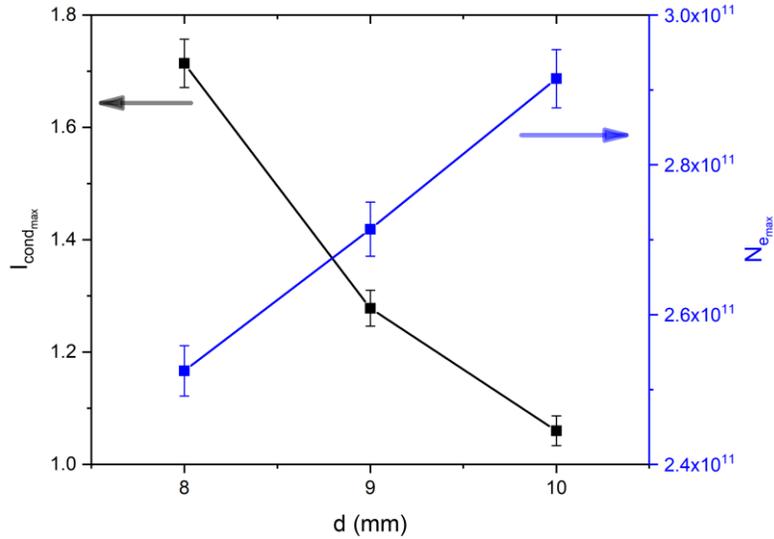

**Figure 16. Dependence of conduction current and peak $N_e$ on gap distance *d* for the corona regime.**

A comparison of the discharge parameters for the two discharge regimes (corona and spark) is shown in Figure 17, including breakdown voltage, peak conduction current, discharge pulse energy, and maximum number of electrons produced (during BD1 for the spark regime). While spark discharges were accompanied with abrupt and loud breakdown events, corona discharge was associated with milder and quieter operation. As shown in Figure 17(a), $V_{br}$ increased monotonically with electrode gap distance, increasing from near 2.5 kV for the spark regime in a 2 mm gap up to nearly 25 kV for the corona regime in 8-10 mm. The corona regime was associated with substantially lower peak of conduction current and discharge pulse energies vs. the spark regime as shown in Figure 17(b) and (c). The discharge pulse energy per unit gap distance was an order of magnitude higher for spark discharges, and was relatively constant within each regime: ~1 mJ/mm for spark and ~0.1 mJ/mm for corona. Figure 17(d) shows that electron production increased for smaller inter-electrode gaps, indicating a higher degree of ionization for shorter gap distances.

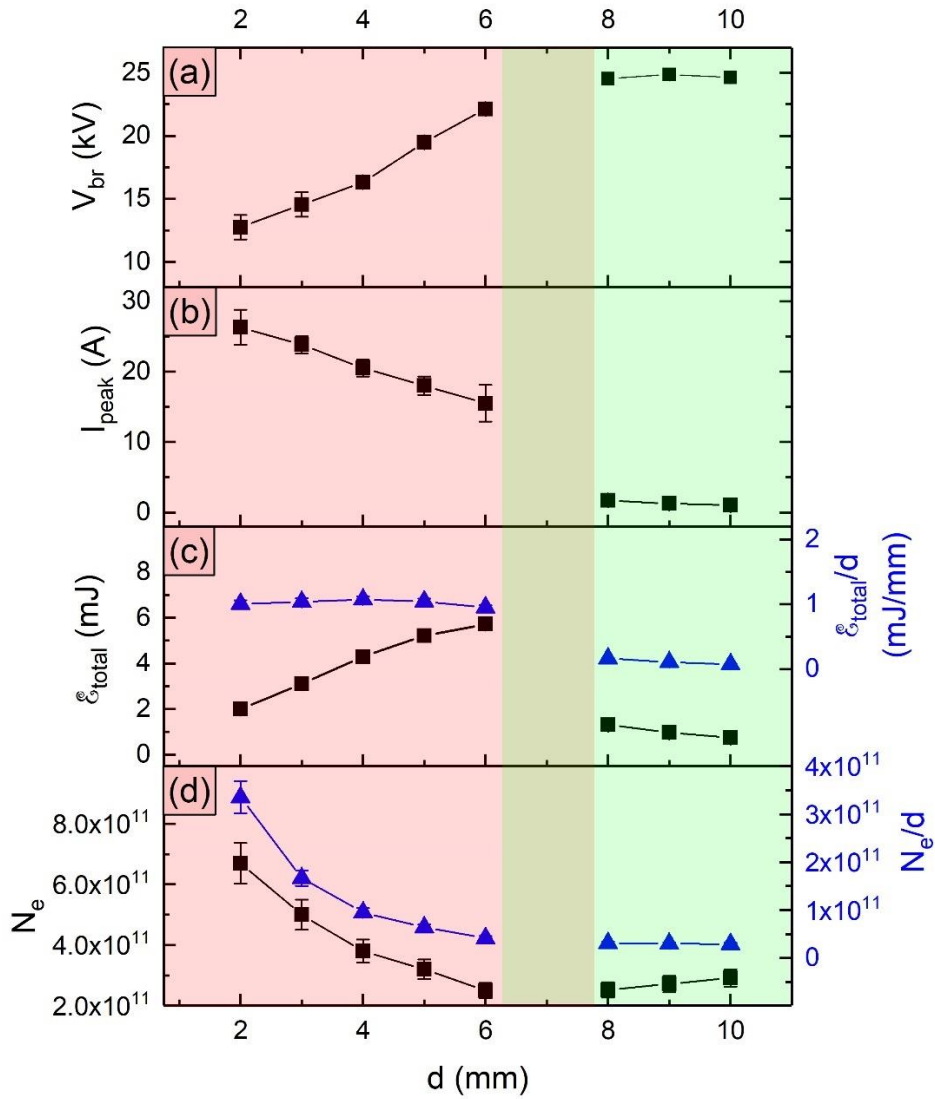

**Figure 17.** Discharge parameters in both discharge regimes vs *d*: (a) breakdown voltage, (b) peak current, (c) total energy input and energy input per gap distance, (d) total electron and electron per gap distance. Spark and corona regimes are indicated by red and green backgrounds, respectively.

# Conclusion

Single-pulse nanosecond high-voltage discharges in pin-to-pin electrode configuration at atmospheric pressure can operate in spark and corona regimes. The spark regime was associated with shorter interelectrode gap sizes (< 6 mm) and larger discharge pulse energy per mm of gap length (0.6-1 mJ/mm) in comparison with the corona regime observed at larger gaps (> 8 mm) and lower discharge pulse energy (~0.1 mJ/mm). Highest electron number density, maximal gas temperature, and peak discharge current are achieved in the spark regime using smaller electrode gaps and higher pulse energy. The characteristics of the spark discharge after breakdown are consistent with the balance between electron production in the discharge volume and electron removal to the anode ensuring conduction of the discharge current, and correspond to reduced electric fields in the range ~100-150 Td. Limitation of the discharge pulse duration to 30-70 ns is necessary in order to prevent transition to the cathodic arc regime. Comparison of experimentally measured and numerically simulated electron decay rates indicate that the initial decay (up to $t \sim$ 1-2 µs after the discharge pulse) is governed by comparable contributions of dissociative recombination and electron attachment to oxygen, while in later stages the decay is expected to be dominated by attachment to oxygen. Based on the measured temporal evolution and recovery of gas number density and temperature, the onset of memory effects and the associated transition to the nanosecond repetitively pulsed (NRP) regime can be expected for pulse repetition frequencies greater than 1 kHz.

# Acknowledgment

Authors would like to thank N. A. Popov, M. N. Shneider, and B. Singh for valuable discussions. This work was supported by the U.S. Department of Energy (Grant No. DE-SC0018156) and partially by the National Science Foundation (Grant No. 1903415).

# Data availability

The data that supports the findings in this study are available from the corresponding author upon reasonable request.

# References

1. Starikovskaia, S. M., Anikin, N. B., Pancheshnyi, S. V., Zatzepin, D. V. & Starikovskii, A. Y. Pulsed breakdown at high overvoltage: Development, propagation and energy branching. *Plasma Sources Sci. Technol.* **10**, 344–355 (2001).

2. Popov, N. A. Fast gas heating in a nitrogen-oxygen discharge plasma: I. Kinetic mechanism. *J. Phys. D. Appl. Phys.* **44**, (2011).

3. Mintoussov, E. I., Pendleton, S. J., Gerbault, F. G., Popov, N. A. & Starikovskaia, S. M. Fast gas heating in nitrogen-oxygen discharge plasma: II. Energy exchange in the afterglow of a volume nanosecond discharge at moderate pressures. *J. Phys. D. Appl. Phys.* **44**, (2011).

4. Kim, W., Godfrey Mungal, M. & Cappelli, M. A. The role of in situ reforming in plasma enhanced ultra lean premixed methane/air flames. *Combust. Flame* **157**, 374–383 (2010).

5. Pham, Q. L. L., Lacoste, D. A. & Laux, C. O. Stabilization of a premixed methane-air flame using nanosecond repetitively pulsed discharges. *IEEE Trans. Plasma Sci.* **39**, 2264–2265 (2011).

6. Pilla, G. L., Lacoste, D. A., Veynante, D. & Laux, C. O. Stabilization of a swirled propane-air flame using a nanosecond repetitively pulsed plasma. *IEEE Trans. Plasma Sci.* **36**, 940–941 (2008).

7. Ombrello, T., Won, S. H., Ju, Y. & Williams, S. Flame propagation enhancement by plasma excitation of oxygen. Part I: Effects of O3. *Combust. Flame* **157**, 1906–1915 (2010).

8. Sun, W., Won, S. H. & Ju, Y. In situ plasma activated low temperature chemistry and subsequent S-curve transition in DME/oxygen/helium mixture. *8th US Natl. Combust. Meet. 2013* **4**, 3384–3391 (2013).

9. Lacoste, D. A., Moeck, J. P., Durox, D., Laux, C. O. & Schuller, T. Effect of nanosecond repetitively pulsed discharges on the dynamics of a swirl-stabilized lean premixed flame. *Proc. ASME Turbo Expo* **1 A**, (2013).

10. Lacoste, D. A., Xu, D. A., Moeck, J. P. & Laux, C. O. Dynamic response of a weakly turbulent lean-premixed flame to nanosecond repetitively pulsed discharges. *Proc. Combust. Inst.* **34**, 3259–3266 (2013).

11. Moeck, J. P. *et al.* Stabilization of a methane-air swirl flame by rotating nanosecond spark discharges. *IEEE Trans. Plasma Sci.* **42**, 2562–2563 (2014).

12. Moeck, J. P., Lacoste, D. A., Laux, C. O. & Paschereit, C. O. *Control of combustion dynamics in a swirl-stabilized combustor with nanosecond repetitively pulsed discharges*. 51st AIAA Aerospace Sciences Meeting including the New Horizons Forum and Aerospace Exposition 2013 (2013) doi:10.2514/6.2013-565.

13. Roupassov, D. V., Nikipelov, A. A., Nudnova, M. M. & Starikovskii, A. Y. Flow separation control by plasma actuator with nanosecond pulse periodic discharge. *46th AIAA Aerosp. Sci. Meet. Exhib.* **47**, (2008).

14. Little, J., Takashima, K., Nishihara, M., Adamovich, I. V. & Samimy, M. Separation control with nanosecond-pulse-driven dielectric barrier discharge plasma actuators. *AIAA J.* **50**, 350–365 (2012).

15. DeBlauw, B., Elliott, G. & Dutton, C. Active control of supersonic base flows with electric arc plasma actuators. *AIAA J.* **52**, 1502–1517 (2014).

16. Samimy, M., Kearney-Fischer, M., Kim, J. H. & Sinha, A. High-speed and high-reynolds-number jet control using localized arc filament plasma actuators. *J. Propuls. Power* **28**, 269–280 (2012).

17. Utkin, U. G. *et al.* Development and use of localized arc filament plasma actuators for high-speed flow control. *J. Phys. D Appl. Phys.* **40**, 605–636 (2007).

18. Samukawa, S. *et al.* The 2012 plasma roadmap. *J. Phys. D. Appl. Phys.* **45**, (2012).

19. Pai, D. Z., Lacoste, D. A. & Laux, C. O. Transitions between corona, glow, and spark regimes of nanosecond repetitively pulsed discharges in air at atmospheric pressure. *J. Appl. Phys.* **107**, (2010).

20. Rusterholtz, D. L., Lacoste, D. A., Stancu, G. D., Pai, D. Z. & Laux, C. O. Ultrafast heating and oxygen dissociation in atmospheric pressure air by nanosecond repetitively pulsed discharges. *J. Phys. D. Appl. Phys.* **46**, (2013).

21. Van Der Horst, R. M., Verreycken, T., Van Veldhuizen, E. M. & Bruggeman, P. J. Time-resolved optical emission spectroscopy of nanosecond pulsed discharges in atmospheric-pressure N 2 and N 2/H 2O mixtures. *J. Phys. D. Appl. Phys.* **45**, (2012).

22. Sainct, F. P., Lacoste, D. A., Laux, C. O., Kirkpatrick, M. J. & Odic, E. *Investigation of water dissociation by Nanosecond Repetitively Pulsed Discharges in superheated steam at atmospheric pressure*. 51st AIAA Aerospace Sciences Meeting including the New Horizons Forum and Aerospace Exposition 2013 (2013) doi:10.2514/6.2013-1190.

23. Zhu, X. M., Walsh, J. L., Chen, W. C. & Pu, Y. K. Measurement of the temporal evolution of electron density in a nanosecond pulsed argon microplasma: Using both Stark broadening and an OES line-ratio method. *J. Phys. D. Appl. Phys.* **45**, (2012).

24. Xu, D. A., Shneider, M. N., Lacoste, D. A. & Laux, C. O. Thermal and hydrodynamic effects of nanosecond discharges in atmospheric pressure air. *J. Phys. D. Appl. Phys.* **47**, (2014).

25. Pai, D. Z., Lacoste, D. A. & Laux, C. O. Nanosecond reprtitively pulsed discharge in air at atmospheric pressure - spark regime. *Plasma Sources Sci. Technol.* **19**, (2010).

26. Wang, X., Stockett, P., Jagannath, R., Bane, S. P. M. & Shashurin, A. Time-resolved measurements of electron density in nanosecond pulsed plasmas using microwave scattering. *Plasma Sources Sci. Technol.* **27**, (2018).

27. Lo, A. *et al.* Streamer-to-spark transition initiated by a nanosecond overvoltage pulsed discharge in air. *Plasma Sources Sci. Technol.* **26**, (2017).

28. Miles, J. *et al.* Time resolved electron density and temperature measurements via Thomson scattering in an atmospheric nanosecond pulsed discharge. *Plasma Sources Sci. Technol.* **29**, (2020).

29. Laux, C. O. *2002 Radiation and Nonequilibrium Collisional-Radiative Models*. (2002).

30. SPECAIR. http://www.specair-radiation.net/.

31. Chelouah, A., Chelouah, A., Chelouah, A. & Chelouah, A. A new method for temperature

evaluation in a nitrogen discharge. *J. Phys. D. Appl. Phys.* **27**, 940–945 (1994).

32. Deng, X. L., Nikiforov, A. Y., Vanraes, P. & Leys, C. Direct current plasma jet at atmospheric pressure operating in nitrogen and air. *J. Appl. Phys.* **113**, (2013).

33. Zhang, Q. Y. *et al.* Determination of vibrational and rotational temperatures in highly constricted nitrogen plasmas by fitting the second positive system of N2 molecules. *AIP Adv.* **5**, (2015).

34. Wang, X. & Shashurin, A. Study of atmospheric pressure plasma jet driven by DC high voltage. *AIAA Aerosp. Sci. Meet. 2018* **122**, (2018).

35. Wang, X. & Shashurin, A. Gas thermometry by optical emission spectroscopy enhanced with probing nanosecond plasma pulse. *AIAA J.* **58**, 3245–3249 (2020).

36. Popa, S. D. Vibrational distributions in a flowing nitrogen glow discharge. *J. Phys. D. Appl. Phys.* **29**, 411–415 (1996).

37. Miles, R. B., Lempert, W. R. & Forkey, J. N. Laser Rayleigh scattering. *Meas. Sci. Technol.* **12**, R33–R51 (2001).

38. Boyda, M., Byun, G. & Lowe, K. T. Investigation of velocity and temperature measurement sensitivities in cross-correlation filtered Rayleigh scattering (CCFRS). *Meas. Sci. Technol.* **30**, (2019).

39. McKenzie, R. L. Progress in laser spectroscopic techniques for aerodynamic measurements: An overview. *AIAA J.* **31**, 465–477 (1993).

40. Zhao, F. Q. & Hiroyasu, H. The applications of laser Rayleigh scattering to combustion diagnostics. *Prog. Energy Combust. Sci.* **19**, 447–485 (1993).

41. Van Gessel, A. F. H., Carbone, E. A. D., Bruggeman, P. J. & Van Der Mullen, J. J. A. M. Laser scattering on an atmospheric pressure plasma jet: Disentangling Rayleigh, Raman and Thomson scattering. *Plasma Sources Sci. Technol.* **21**, 015003 (2012).

42. Limbach, C. M., Dumitrache, C. & Yalin, A. P. *Laser light scattering from equilibrium, high temperature gases: Limitations on rayleigh scattering thermometry*. *47th AIAA Plasmadynamics and Lasers Conference* (2016) doi:10.2514/6.2016-3381.

43. Wang, X., Patel, A. R. & Shashurin, A. Combined microwave and laser Rayleigh scattering diagnostics for pin-to-pin nanosecond discharges. *J. Appl. Phys* **129**, (2021).

44. Shneider, M. N. & Miles, R. B. Microwave diagnostics of small plasma objects. *J. Appl. Phys.* **98**, 033301 (2005).

45. Sharma, A. *et al.* Counting the electrons in a multiphoton ionization by elastic scattering of microwaves. *Sci. Rep.* **8**, (2017).

46. Shashurin, A., Shneider, M. N., Dogariu, A., Miles, R. B. & Keidar, M. Temporary-resolved measurement of electron density in small atmospheric plasmas. *Appl. Phys. Lett.* **96**, 171502 (2010).

47. Sharma, A. *et al.* Diagnostics of CO concentration in gaseous mixtures at elevated pressures by resonance enhanced multi-photon ionization and microwave scattering. *J. Appl. Phys.* **128**, (2020).

48. Raizer, Y. P. *Gas Discharge Physics*. (Springer-Verlag Berlin Heidelberg, 1987).

49. Roettgen, A., Shkurenkov, I., Simeni Simeni, M., Adamovich, I. V. & Lempert, W. R. Time-resolved electron temperature and electron density measurements in a nanosecond pulse filament discharge in H2-He and O2-He mixtures. *Plasma Sources Sci. Technol.* **25**, (2016).

50. Shao, T. *et al.* Spark discharge formation in an inhomogeneous electric field under conditions of runaway electron generation. *J. Appl. Phys.* **111**, (2012).

51. Tarasenko, V. F. *et al.* Generation of supershort avalanche electron beams and formation of diffuse discharges in different gases at high pressure. *Plasma Devices Oper.* **16**, 267–298 (2008).

52. Beilis, I. *Plasma and Spot Phenomena in Electrical Arcs*. (Spinger International Publishing, 2020).

53. Limbach, C. M. Characterization of nanosecond, femtosecond and dual pulse laser energy deposition in air for flow control and diagnostic applications. (Princeton University, 2015).

54. Itikawa, Y. Cross section for electron collisions with nitrogen molecules. *J. Phys. Chem. Ref. Data* **35**, (2006).

55. Itikawa, Y. & Ichimura, A. Cross Sections for Collisions of Electrons and Photons with Atomic Oxygen. *J. Phys. Chem. Ref. Data* **19**, 637–651 (1990).

56. Hake, R. D. & Phelps, A. V. Momentum-transfer and inelastic-collision cross sections for electrons in O2, CO, and CO2. *Phys. Rev.* **158**, 70–84 (1967).

57. Popov, N. A. Pulsed nanosecond discharge in air at high specific deposited energy: Fast gas heating and active particle production. *Plasma Sources Sci. Technol.* **25**, (2016).

58. Dogariu, A., Shneider, M. N. & Miles, R. B. Versatile radar measurement of the electron loss rate in air. *Appl. Phys. Lett.* **103**, (2013).

59. Kossyi, I. A., Kostinsky, A. Y., Matveyev, A. A. & Silakov, V. P. Kinetic scheme of the non-equilibrium discharge in nitrogen-oxygen mixtures. *Plasma Sources Sci. Technol.* **1**, 207–220 (1992).

60. Dutton, J. A survey of electron swarm data. *J. Phys. Chem. Ref. Data* **4**, 577–856 (1975).

61. Stepanyan, S. *et al.* Spatial evolution of the plasma kernel produced by nanosecond discharges in air. *J. Phys. D. Appl. Phys.* **52**, (2019).

62. Singh, B., Rajendran, L. K., Vlachos, P. P. & Bane, S. P. M. Two Regime Cooling in Flow Induced by a Spark Discharge. *Phys. Rev. Fluids* **5**, 14501 (2019).

63. Singh, B., Rajendran, L. K., Zhang, J., Vlachos, P. P. & Bane, S. P. M. Vortex rings drive entrainment and cooling in flow induced by a spark discharge. *Phys. Rev. Fluids* **5**, (2020).